\documentclass[aps,prl,preprint,groupedaddress,showpacs]{revtex4}
\usepackage{epsfig}
\pdfoutput=1
\usepackage[colorlinks=true,pagebackref=true]{hyperref}
\hypersetup{
    pdftitle={Spin-Orbit Coupling of Conduction Electrons in Magnetization Switching},
    pdfauthor={Ioan Tudosa},
    pdfsubject={},
    pdfkeywords={magnetism magnetization switching magneto-transport}
    }

\begin{document}

\title{Spin-Orbit Coupling of Conduction Electrons in Magnetization Switching}

\author{Ioan Tudosa}
\noaffiliation

\date{\today}

\begin{abstract}
Strong magnetic field pulses associated with a relativistic electron bunch can imprint switching patterns in magnetic thin films that have uniaxial in-plane anisotropy. In experiments with Fe and FeCo alloy films the pattern shape reveals an additional torque acting on magnetization during the short (in the 100fs time scale) magnetic field pulse. The magnitude of the torque is as high as 15\% of the torque from the magnetic field. The torque symmetry is that of a uniaxial anisotropy along the direction of the eddy current screening the magnetic field. Spin-orbit interaction acting on the conduction electrons can produce such a torque with the required symmetry and magnitude. The same interaction causes the anomalous Hall current to be spin-polarized, exerting a back reaction on magnetization direction. Such a mechanism may be at work in all-optical laser switching of magnetic materials.
\end{abstract}


\maketitle

\section{}

Magnetization switching and dynamics has been tested down to femtosecond time scale with the help of lasers and relativistic short electron pulses \cite{Back:1998,Tudosa:2004,Stanciu:2007,Ostler:2012}. While the exact laser induced switching mechanism is still hotly debated \cite{Ellis:2016}, the dynamics under the short magnetic field pulse associated with a relativistic electron bunch was understood as a precession due to magnetic field torque acting on magnetization \cite{Back:1998}. Recent results of experiments \cite{Tudosa:2017} with Fe and FeCo alloy thin films at SLAC National Accelerator Laboratory show that there is an additional torque from the spin-orbit coupling of conduction electrons in the eddy currents screening the rapidly varying magnetic field. This additional torque can be as large as 15\% of the magnetic field torque and has a different symmetry that makes its identification easier.

\begin{figure}
\includegraphics[]{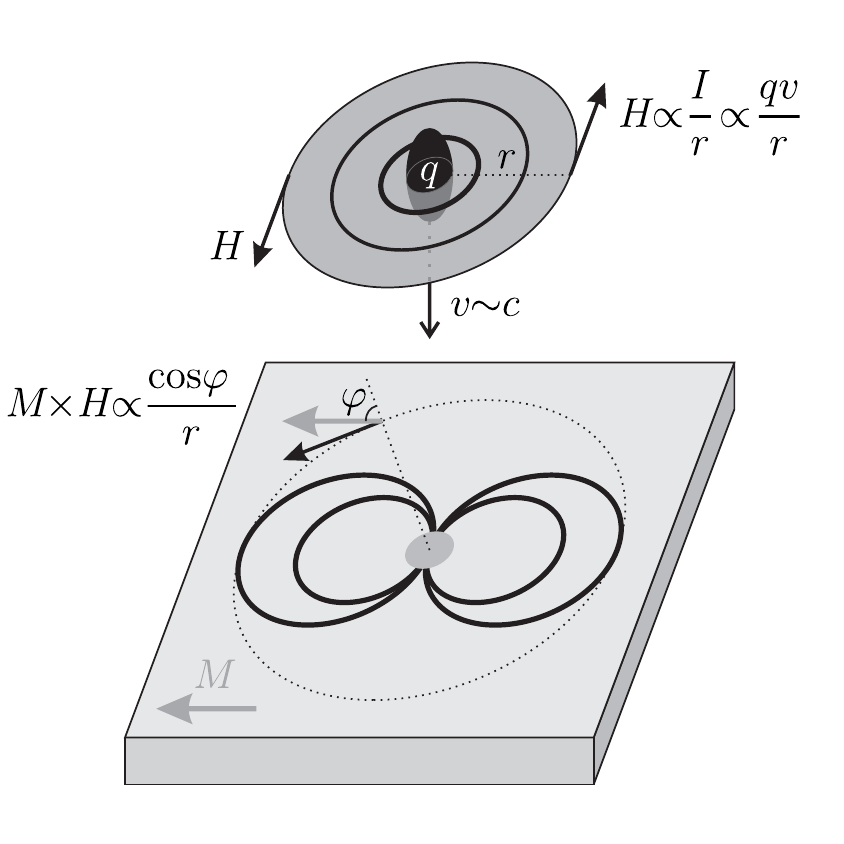}
\caption{\label{experiment} Schematic of the magnetic switching experiment using a short relativistic electron bunch with energy around 20 GeV. The speed of the bunch is very close to that of light and the magnetic and electric fields lie in a plane perpendicular to the propagation direction. Electric field lines are radial while the magnetic field lines are circular, with a 1/r dependence and direction \(\vec{E}=\vec{v}\times\vec{B}\) (a configuration similar to that of a radiating electromagnetic field). Magnetic fields in the range of 10-100 Tesla can be thus generated. After the electron bunch has passed through, the magnetic pattern encodes the magnetization switching dynamics in response to the torque exerted by the magnetic field of the bunch. The pattern will follow the lines of constant torque \(\frac{\cos\varphi}{r}=const\) in the form of a figure 8 with perfect circular lobes. In metallic samples the electric field is screened and plays no role in generating the pattern.}
\end{figure}

In a typical setup an electron bunch is accelerated and slammed into a sample. As the relativistic electron bunch passes through a uniformly magnetized magnetic thin film sample, its magnetic field will excite the film magnetization and imprint a magnetic pattern, as illustrated in Fig. \ref{experiment}. This pattern can reveal details about the magnetization dynamics. Our samples are initially magnetized in-plane along an easy-axis given by the in-plane uniaxial crystalline anisotropy. The magnetic field makes the magnetization come out of the sample plane and after the field pulse is gone the magnetization will precess around the demagnetization field, eventually getting damped out and settling into an easy axis direction, as shown in Fig. \ref{Damping}. The pattern looks like a figure 8, where boundaries are lines of constant torque acting during the field pulse, \(\frac{\cos\varphi}{r}=const\), with \(\varphi\) being angle between magnetization and radial distance \(r\) from the impact point. The circular magnetic field has a \(\frac{1}{r}\) dependence and exerts a torque proportional to \(\cos\varphi\) on the magnetization.

\begin{figure}
\includegraphics[]{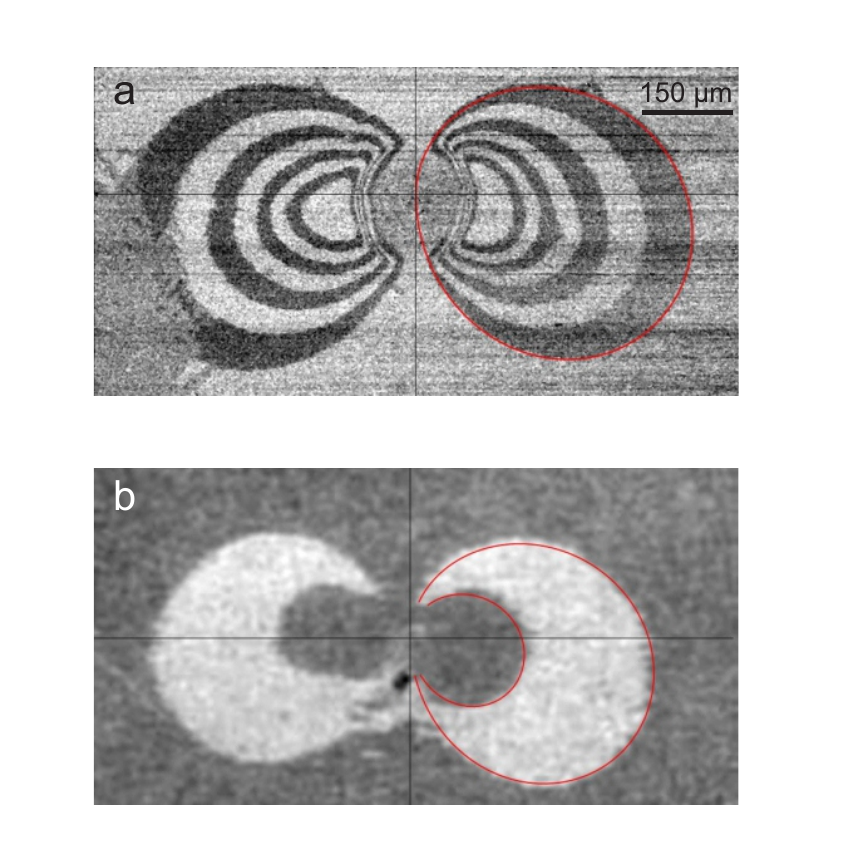}
\caption{\label{SampleImages} Magnetic contrast images taken with a scanning electron microscope with polarization analysis. Black and white colors represent opposite directions of magnetization along the horizontal axis. The samples were initially uniformly magnetized and exposed to an electron bunch as in Fig. \ref{experiment}. Red lines represent best fit of the switching boundaries. (a) The magnetic pattern on a W(110)/Fe(60ML)/Ag(4nm) sample. The equation for the fit in polar coordinates is \(\frac{\cos\varphi-\kappa\sin 2\varphi}{r}=const\), where \(\kappa=0.15\) is a constant expressing the ratio of the additional torque to the magnetic field torque. A damping of 0.013 is extracted from the position of the rings as shown in Fig \ref{Damping}. (b) The magnetic pattern on a MgO(110)/Cr(30nm)/Co70Fe30(100nm)/Pt(1.5nm) sample. A similar ratio of \(\kappa=0.15\) for additional torque fits the pattern very well. A damping value of 0.06 is extracted from the position of rings.}
\end{figure}

\begin{figure}
\includegraphics[]{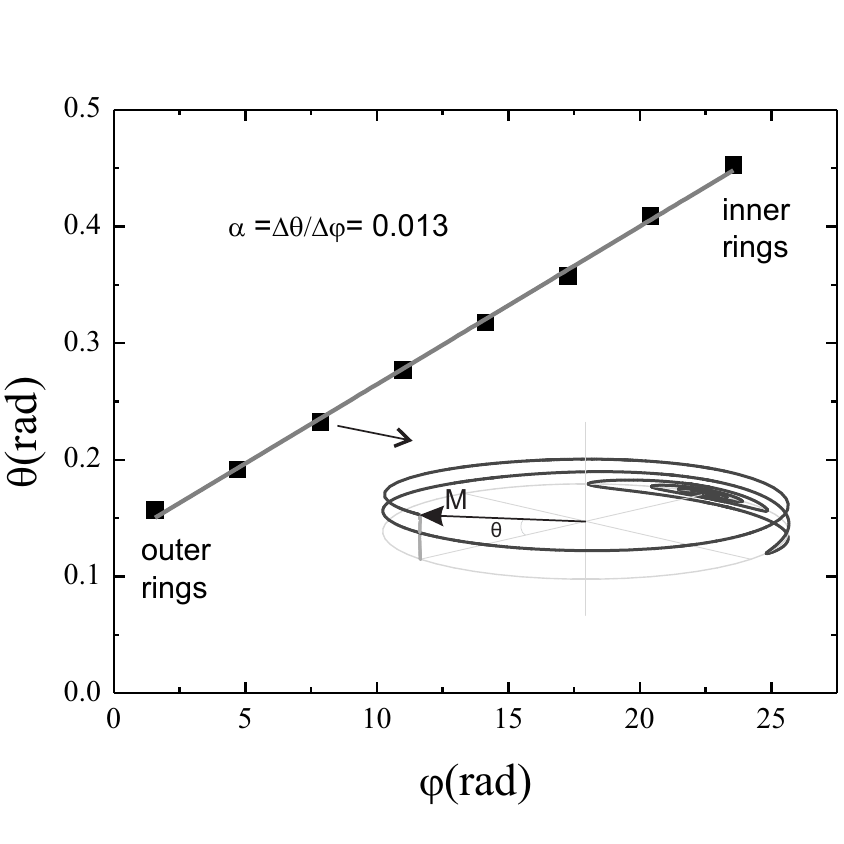}
\caption{\label{Damping} Damping can be extracted from fitting of the in plane precession angle \(\varphi\) versus magnetization out of plane deviation angle \(\theta\), shown here for a W(110)/Fe(60ML)/Ag(4nm) sample. The inset illustrates the magnetization dynamics. The gray initial part of the trace is the precession out of plane during the short field pulse. The damping process brings the magnetization back into the plane, decreasing the angle \(\theta\),increasing \(\varphi\) and eventually snapping into one of the two easy directions of the uniaxial in-plane anisotropy. The black squares represent the location of the switching boundaries (rings).}
\end{figure}

Fig. \ref{SampleImages} shows the main results of experiments with very short electron bunches having \(N_e \approx 2 \times 10^{10}\) electrons in a pulse with Gaussian distribution \(\sigma_{x,y,z} \approx 30\times30\times30\mu m\), moving close to the speed of light at 20 GeV energy. One notices right away that the pattern is not a figure 8 with perfect circular lobes. An additional torque acts on the magnetization besides the torque from magnetic field of the electron bunch. This additional torque has an angular dependence of $\sin2\varphi$, where $\varphi$ is the angle between radial direction from impact point and magnetization. The presence of this effect in samples with various thicknesses and layer materials (shown in Table \ref{Comparison}) seems to favor an intrinsic volume effect, not an interface phenomenon.

The magnetization dynamics can be modeled by Landau-Lifshitz-Gilbert equation \ref{eq:LLG} with the addition of a spin torque due to the non-equilibrium spin density of conduction electrons \(\Delta\vec{m}\). This spin density exerts a torque \cite{Manchon:2012} on the bulk magnetization \(\vec{M}\) (with spin \(S\) and saturation \(M_s\)) through the s-d exchange interaction with Hamiltonian \(H_{sd}=-J_{sd}\vec{S}\cdot\vec{s}\), where \(J_{sd}\) is the exchange energy, \(\vec{s}\) is the spin of the conduction electron. This torque has also other acronyms such as NEXI (non equilibrium exchange interaction) torque \cite{Stohr:2006}.

\begin{equation}
 \label{eq:LLG}
 \frac{d\vec{M}}{dt} = -\gamma\vec{M}\times\vec{H} + \frac{\alpha}{M_s}\vec{M}\times\frac{d\vec{M}}{dt} - \frac{J_{sd}S}{\hbar M_s}\vec{M}\times\Delta\vec{m}.
\end{equation}

The effective damping can be estimated from the precessions around the demagnetization field for each ring boundary. As magnetization precesses around the demagnetization field (expressed by variation of angle \(\varphi\)) the damping brings it slowly into the plane (expressed by angle \(\theta\) between magnetization and sample plane). The damping is then simply the derivative \(\alpha=\frac{d\theta}{d\varphi}\), as shown in Fig. \ref{Damping}. It is constant over the whole range of magnetic field intensities and magnetization deviation angle, in contrast with a previous paper where variable damping was employed \cite{Sara:2009} in the interpretation.

In metallic samples there is a strong radial eddy current due to screening of the magnetic field of the electron bunch. The skin depth for 100fs bunch duration is on the order of 100nm and assuming a typical field of 10T gets screened over this distance the eddy current density (H/skin depth) is on the order of $10^{10} A/cm^2$, a pretty large current. Assuming the additional torque to be 0.15 of the external field torque as fitted in Fig. \ref{SampleImages}, the additional torque to current ratio is about $1.5mT / 10^{7} A/cm^2$ in the range of torque values measured experimentally \cite{Garello:2013} for currents passing through heavy metal/ferromagnetic heterostructures. The fitting gives at least values comparable to literature.

The optical constants \cite{Ordal:1988,Wooten:1972} for iron at 10 THz (the 100fs time scale) show a dielectric constant \(\epsilon=5000\), making the electric field inside very small \(E_{metal}=E_{vacuum}/\epsilon\). The outside electric field is screened at the surface \cite{Jackson:1998} of the metal, while only the magnetic field penetrates inside with a sizable strength. The small electric field in the metal is driving the eddy currents that screen the magnetic field (via Faraday induction law). This is 3-4 orders of magnitude smaller than the electric field associated with the electron bunch in vacuum. There is simply not a sizable electric field inside the metal to induce bulk magnetic anisotropy by distorting the electronic orbitals as assumed elsewhere \cite{Sara:2009}. The metallic capping and capping layer also prevent any electric field induced surface anisotropy.

\begin{table}%[H] add [H] placement to break table across pages
\caption{\label{Comparison} Illustration of differences and similarities between two samples. The capping and the substrate do not matter much and neither the thickness. Damping also does not play a significant role.}
\begin{ruledtabular}
\begin{tabular}{lcr}
\bf{Differences} & Fe/W(110) & CoFe/MgO(110) \\
\hline
Material & Fe & CoFe alloy \\
Substrate & metallic & insulator \\
Capping & Ag & Pt \\
Thickness & ~10nm & 100nm \\
Damping & 0.013 & 0.06 \\
\hline\hline
\bf{Similarities} & Fe/W(110) & CoFe/MgO(110) \\
\hline
Epitaxial layer & (110) & (110) \\
Anisotropy field & 1kOe & 1kOe \\
%Resistivity (\(\mathrm{\mu\Omega cm}\)) & 10 & 7 \\
Saturation Magnetization & 2.2T & 2T \\
Skin depth (at 10THz) & 50nm & 42nm \\
\end{tabular}
\end{ruledtabular}
\end{table}

\begin{figure}
\includegraphics[]{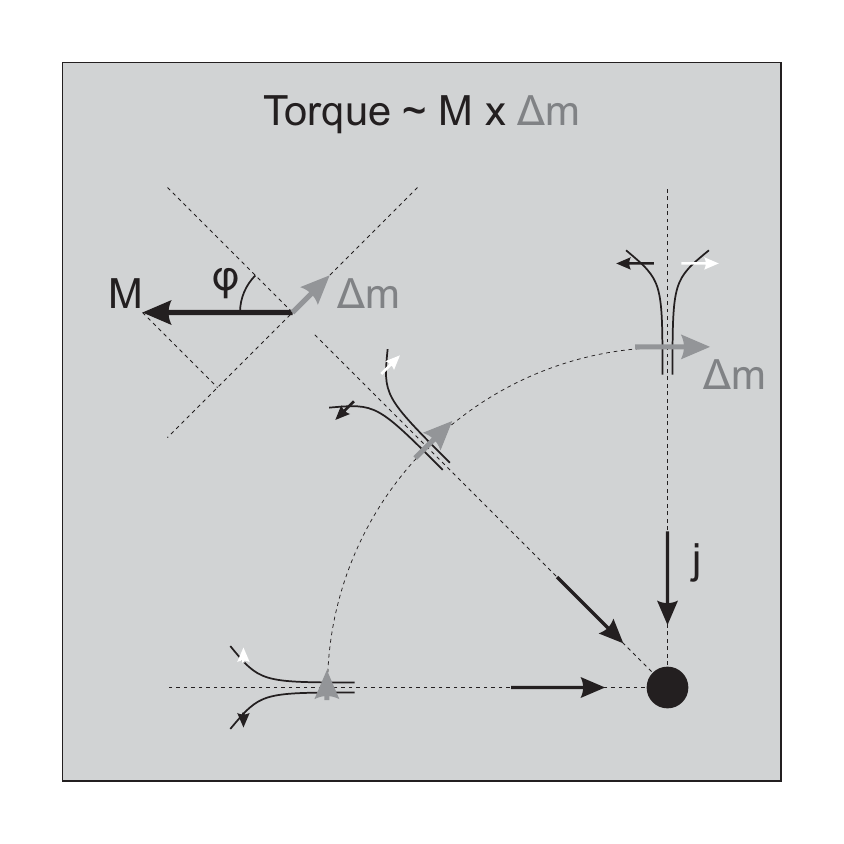}
\caption{\label{SOtorques} The current \(j\) generates a torque on magnetization \(M\) via the additional polarization of conduction electrons \(\Delta m\). This polarization is generated through the spin-orbit coupling effects (similar to anomalous Hall effect and anisotropic magnetoresistance) and is perpendicular to the current. Only conduction electrons with spins that are perpendicular to their velocity (shown here as black and white arrows) will feel these effects, and their number is proportional to the magnetization component perpendicular to current. The additional polarization exerts a torque proportional to \(\vec{M}\times\Delta \vec{m}\) on magnetization, through the s-d exchange interaction. If the angle between magnetization and current directions is \(\varphi\), then the torque will have a \(\sin\varphi\cdot\cos\varphi\propto\sin 2\varphi\) dependence seen in figure \ref{SampleImages}. }
\end{figure}

The spin-orbit coupling of conduction electrons in ferromagnetic materials generate effects like anisotropic magneto-resistance (AMR) and anomalous Hall effect (AHE) \cite{Nagaosa:2010,Kokado:2012}. These effects relate electrical measurements to orientation of the magnetization. Less explored is what happens to the magnetization when the current is high. Is there a back reaction, a torque, acting on magnetization? The spin-orbit coupling involved in AMR and AHE make electrons scatter \cite{Nagaosa:2010} to one side or another depending on their spin, generating a separation of spins and leading to a spin current and spin accumulation. On general arguments the additional polarization generated in conduction electrons by a spin-orbit interaction can be only perpendicular to the direction of the motion, that is, the current direction as seen in Fig. \ref{SOtorques}. A spin parallel to the velocity direction does not feel the spin-orbit interaction.

The spin-orbit interaction for a conduction electron has the Hamiltonian of the form \(H_{SO}\propto(s\cdot \nabla V)\times p\), where \(p\) is momentum of the electron moving through a potential \(V\). This implies that only the spin component perpendicular to the momentum \(s_{\perp}\propto\sin\varphi\) contributes to this interaction, in our case the spin perpendicular to the current direction (easily seen by doing a relativistic transformation of the electromagnetic fields in the frame of reference of the moving electron). This interaction separates spin up and spin down (along the direction of \(\vec{s}_{\perp}\)) contributing to a non-equilibrium accumulation of \(\Delta \vec{s}_{\perp}\). The s-d exchange energy associated with this spin accumulation is \(E=-J_{sd}\vec{S}\cdot\Delta \vec{s}_{\perp}\propto\sin^2\varphi\) and so the exchange torque has an angular dependence \(T=\frac{\partial E}{\partial\varphi}\propto\sin2\varphi\).

Only the spin component perpendicular to the current feels the spin-orbit interaction leading to a spin accumulation proportional to \(\sin\varphi\), where \(\varphi\) is the angle between current and magnetization. This non-equilibrium polarization of conduction electrons acts on magnetization with a torque via the s-d exchange interaction in equation \ref{eq:LLG}, acquiring an additional factor of \(\cos\varphi\) and the overall angular dependence of \(\sin 2\varphi\). The lines of constant total torque have now the expression \(\frac{\cos\varphi-\kappa\sin 2\varphi}{r}=const\), where \(\kappa\) is a constant expressing the ratio of the additional spin-orbit torque to the magnetic field torque.

At 0 and 90 degrees the torque is zero while it is maximum at 45 degrees, a fact that is reflected in the shape of the pattern in figure \ref{SampleImages} where lobes are distorted towards 45 degrees. At 0 degree the conduction electron spin is parallel to the current, so few electrons suffer spin-orbit deflection, while at 90 degrees many spins electron experience spin-orbit deflection but the s-d exchange torque is zero because of parallel orientation of spin accumulation and magnetization.

Regardless of the details of spin-orbit interaction, a spin up will be deflected to one side and a spin down will be deflected to the other side as they are moving with the charge current. The origin of the spin-orbit coupling may be the same as in AHE in ferromagnetic materials: intrinsic deflection related to the Berry phase curvature and inter-band coherence, side jump and skew scattering on impurities \cite{Nagaosa:2010}. As soon as spins separate a spin accumulation appears. The separation of spins will create a spin current with direction and polarization perpendicular to the charge current. In places where this polarization is not parallel with the magnetization there will be spin scattering flipping some spin-up into spin-down. This scattering is very efficient in ferromagnets on the order of fs time scale and converts a spin current into a spin accumulation that will exert a torque on the magnetization. The anomalous Hall current is in fact a spin-polarized current (see Fig. 3 of reference \cite{Nagaosa:2010}).

Here we estimate the size of back reaction on the magnetization. The field pulse and material parameters determine the magnitude of the torque. We start with the screening current density \(j\) that has an exponential decay with distance from the surface of the metal as the magnetic field diffuses into the metal \cite{Knoepfel:2000}. Assuming most of the magnetic field is screened by the eddy current in the skin depth, the average current density is roughly \(j\propto\frac{H}{\lambda}\), where H is the amplitude of the external field pulse and \(\lambda\) is its skin depth in the material. As electrons get deflected perpendicular to \(j\) due to spin-orbit interaction, the anomalous Hall current is \(j_{\perp}=\frac{\sigma_{\perp}}{\sigma_\parallel}j\), where \(\sigma_\parallel\) is the regular (longitudinal) conductivity associated with screening current direction, while \(\sigma_\perp\) is the anomalous Hall conductivity associated with the perpendicular deviation from that direction \cite{Nagaosa:2010}. This perpendicular current is totally spin polarized and the spin current density is \(j_s\propto\frac{j_{\perp}}{e P}\), where \(P\) is the polarization of the conduction electrons. All deflected electrons contribute to the spin current but only the spin asymmetry fraction contributes to the charge current, because spin up is deflected to one side and spin down to other side and both have the same negative charge. Since a current density is a density times a velocity, the spin and magnetic moment accumulation (volume density) generated through spin-flip scattering is roughly \(\Delta m\propto\frac{j_s}{v_{F}}\mu_B\), where \(v_{F}\) is Fermi velocity and \(\mu_B\) is the Bohr magneton. The torque from this accumulation is \(T=\frac{J_{sd}S}{\hbar M_s}\Delta \vec{m}\times \vec{M}\) and its magnitude scales like \(T\propto\frac{\mu_B}{\hbar e P}\frac{\sigma_\perp}{\sigma_\parallel}\frac{J_{sd}S}{\lambda v_{F}M_s}H\). The ratio factor between this torque and that of the magnetic field (\(\propto\gamma H\)) is
\begin{equation}
 \label{eq:kappa}
 \kappa=\frac{J_{sd}S}{g e P \lambda v_{F} M_s}\frac{\sigma_\perp}{\sigma_\parallel},
\end{equation}
which for some representative values for iron \cite{Himpsel:1998,Mamy:2002,Nagaosa:2010} \(g=2.09,P=0.4,S=1,J_{sd}=0.7eV,v_F=0.5\times10^6m/s,\frac{\sigma_\perp}{\sigma_\parallel}=0.01,M_s=2.2T,\lambda=50nm\) gives a value of 0.15. Lower saturation magnetization, short field pulses and small skin depth make this torque stronger.

In the case a radiation pulse from an optical laser the skin depth is small but its magnetic field will oscillate in sign and/or direction, in contrast with a unipolar field from an electron bunch. When the magnetic field changes sign over a period the net effect of the magnetic field torque is zero. The accumulation of conduction electrons \(\Delta \vec{m}\) from spin-orbit coupling scattering does not change sign with reversing the current direction, and so the torque acting on magnetization will also not change sign (\(\propto \vec{M} \times \Delta \vec{m}\)). The spin orbit coupling imprints a curvature on the trajectory of the conduction electron that depends only on its spin and not on the direction of motion along the trajectory. It is this bending that manifests itself in the appearance of an additional torque.

\begin{figure}
\includegraphics[]{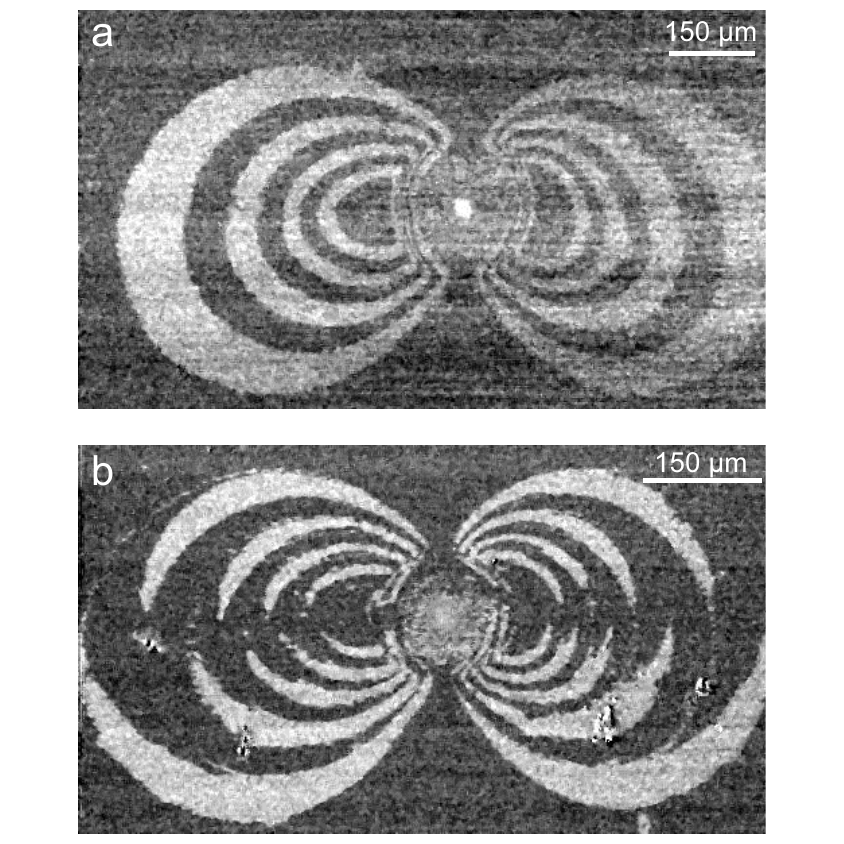}
\caption{\label{FirstSecondShot} The distortion of the patterns is not an artifact of imaging. If the patterns were perfect circles (only magnetic field torque present) the second bunch in the same location would erase the pattern from the first bunch, restoring a uniform magnetization state. This second bunch pattern can be thought as a combination of single bunch patterns, one mirror image of the other due to opposite initial magnetization. Since the single shot pattern is not symmetric around the direction of magnetization, the white regions in the bottom panel remaining white on any later shots. On passing bunches repeatedly through the same spot, the patterns will cycle between top and bottom panel.}
\end{figure}

When we consider only the magnetic field torque a second pulse will deterministically reverse the magnetization and erase the pattern created by the first pulse. The Fig. \ref{FirstSecondShot} shows that this reversible symmetry is broken by the addition of the spin-orbit torque. While the magnetic field torque has a p-like symmetry (\(\propto M\cos\varphi\)), the spin-orbit torque has a d-like symmetry (\(\propto M\cos2\varphi\)). A p-like symmetry is preserved under a 180 degrees rotation (a switching of magnetization) but not a d-like symmetry. Some locations in the sample will get stuck in the d-like symmetry and not be accessible to the cycles of writing and erasure. A simple procedure to obtain the second pulse pattern is as follows: consider the single pulse pattern tied to the initial  magnetization direction; rotate the magnetization and pattern by 180 degrees; take a XOR ($\oplus$) operation between the two images (areas of overlap with different colors will not be erased).

For this torque symmetry to occur it is essential for the conduction carriers to have a fixed direction of polarization, given by magnetization in ferromagnetic materials. When there is no polarization, as in Pt or W used for substrate or capping layers, the spin Hall effect causes a spin current to flow into the ferromagnetic layer and the symmetry of the torque contains only a factor of \(\cos\varphi\), same as that of the external magnetic field. The injected spins have the same orientation as the external magnetic field, and their exchange torque have a similar symmetry. It is unlikely for spin polarized conduction electrons from the ferromagnetic layer to flow into the adjacent heavy metal layer, feel the spin-orbit coupling and come back into ferromagnetic layer to exert a torque on it \(\propto\sin2\varphi\). In any case this will be a surface effect and not a volume effect as required by similar results on thicker samples.

The torque is proportional to current across the whole pattern so nonlinear effects are not at play. The energy associated with this torque (integral over angle) is proportional to \(\sin^2\varphi\), the same as for a uniaxial anisotropy, as alluded in \cite{Gambardella:2011}. In essence the direction of current provides an easy axis for magnetization and for large enough currents this anisotropy can be greater than crystalline anisotropy. Such an effect may be useful in controlling magnetization switching even in homogenous materials and structures.

\begin{figure}
\includegraphics[]{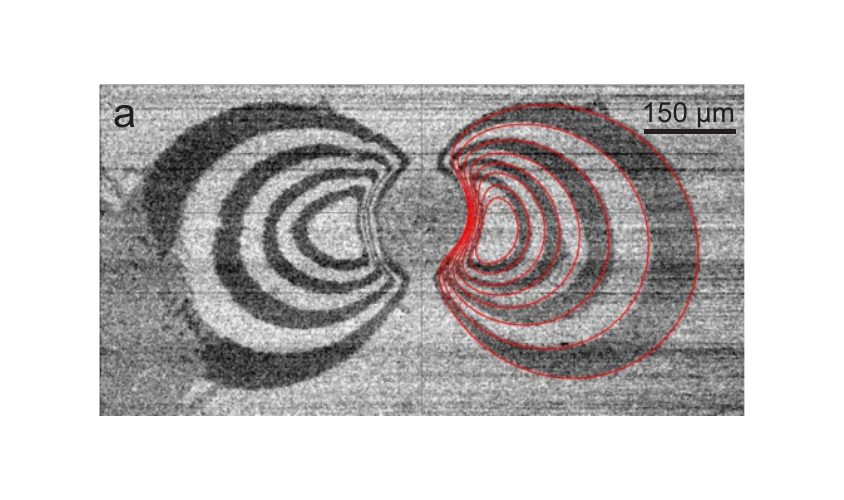}
\caption{\label{FeWheating} The electron bunch with a Gaussian distribution deposits some energy in the sample and by heating above Curie temperature it erases the magnetic pattern near the center and distorts it nearby. A good fit is obtained by employing a multiplicative factor \(1-Ae^{-\frac{r^2}{2r_c^2}}\), with \(A\) and \(r_c\) as fitting parameters. This reproduces the main features because saturation magnetization \(M_{sat}\) and implicitly the demagnetizing field become smaller as temperature increases towards Curie temperature.}
\end{figure}

The pulse duration does not have an influence on the pattern and Gaussian pulses with standard deviation of 100fs, 200 fs and even 4.4ps show similar pattern shapes. The energy deposited by the relativistic electron bunch in the thin film sample can in some cases heat the sample above the Curie temperature erasing the magnetic pattern near the center. The heating is stronger in metallic substrates made of high Z materials and has a range similar to the spatial extent of the bunch (a Gaussian profile), as shown in Fig. \ref{FeWheating}. Substrates with low Z often show little to no damage in the center as in Fig. \ref{SampleImages}b.

To summarize, a magnetic field pulse exerts an additional torque on magnetization in metallic ferromagnetic samples, a consequence of the spin-orbit coupling of the polarized conduction electrons in the eddy current screening the field. A spin-orbit coupling separates spins perpendicular to the motion direction (electric current) and this separation creates a non-equilibrium polarization of conduction electrons that acts back on the total magnetization. This effect does not depend on the sign of the current and is equivalent to an uniaxial anisotropy along the current direction.

\bibliography{pattern}

\end{document}